\documentclass[reprint, amsmath,amssymb,aps,twocolumn]{revtex4}

\usepackage{graphicx}
\usepackage{dcolumn}
\usepackage{bm}
\usepackage{aas_macros}
\usepackage{amsmath}
\usepackage{comment}

\begin{document}

\title{Pair luminosity and cooling of newborn strange star: unpaired quarks.}

\author{Mikalai Prakapenia$^{1,2}$}
\author{Gregory Vereshchagin$^{1,3,4,5}$}
\affiliation{$^{1}$ICRANet-Minsk, Institute of Physics, National Academy of Sciences of Belarus\\
220072 Nezalezhnasci Av. 68-2, Minsk, Belarus}
\affiliation{$^{2}$Department of Theoretical Physics and Astrophysics, Belarusian State University, Nezalezhnasci Av. 4, 220030 Minsk, Belarus}
\affiliation{$^{3}$ICRANet, 65122 Piazza della Repubblica, 10, Pescara, Italy}
\affiliation{$^{4}$ ICRA, Dipartimento di Fisica, Sapienza Universit\`a di Roma, Piazzale Aldo Moro 5, I-00185 Rome, Italy}
\affiliation{$^{5}$INAF -- Istituto di Astrofisica e Planetologia Spaziali, 00133 Via del Fosso del Cavaliere, 100, Rome, Italy}
\date{\today}

\begin{abstract}
It was shown that pair luminosity of the newborn strange star with temperature of $10^{11}$ K may be as high as $L_\pm\simeq 10^{52}$ erg/s. The question remains: can a strange star maintain such a high surface temperature for a long time? To answer this question we studied thermal evolution of newborn strange star  made of unpaired quarks taking into account its thermal conductivity and neutrino emission by the URCA processes $d\rightarrow u + e + \bar\nu_e$ and $u+e\rightarrow d+\nu_e$. Our results show that extremely high luminosity due to the Schwinger process and insufficient thermal conductivity of quarks leads to development of steep temperature gradient at the surface of strange star. As a result, the temperature at the surface and hence its luminosity decreases, reaching $10^{43}$ erg/s already at $10^2$ seconds. This result holds even in the presence of neutrinosphere.
\end{abstract}

\maketitle

\section{Introduction}

The strange quark matter (SQM) hypothesis \cite{1971PhRvD...4.1601B, 1984PhRvD..30..272W} states that quark matter containing approximately equal numbers of up, down, and strange quarks is the true ground state of strongly interacting matter. From the SQM hypothesis a new class of astrophysical compact objects is predicted: quark stars or strange stars \cite{1986ApJ...310..261A, 2005PrPNP..54..193W} made up entirely of quark matter.

Strange star can be formed as a result of phase transition in a neutron star \cite{2003ApJ...586.1250B,2004A&A...416..991A,2013PhRvD..87j3007P} or in a binary merger \cite{2022PhLB..83337388S}. In any case, a newborn strange star is very hot. The temperature in the stellar interior can be as high as $10^{11}$ K, see \cite{chengDai2001,1991ApJ...375..209H,2007ApJ...659.1519D}.  Powerful neutrino emission blows away the envelope of hadron matter, so the quark surface is bare and remains so as long as the temperature is higher than $\simeq10^7$ K \cite{Usov_1997, Usov_2001}.

Neutron combustion was shown to produce a huge energy release of up to $10^{53}$ ergs. Thus, strange stars are considered as possible progenitors of explosive phenomena such as supernovae and gamma-ray bursts \cite{OLINTO198771, PhysRevD.111.063040,2003ApJ...586.1250B,Haensel:2006kkg} or fast radio bursts \cite{2021Innov...200152G}. At the same time, a newborn strange star can be a source of powerful pair emission with luminosity of about $10^{52}$ erg/s, as was shown in \cite{1998PhRvL..80..230U}. The purpose of this paper is to consider this scenario in more detail.

One of the unique features of strange stars is the existence of supercritical electric field in the thin (few hundred fm) layer on its surface \cite{1986ApJ...310..261A}. This layer is called electrosphere \cite{2005ApJ...620..915U}. The strength of electric field $E$ there exceeds the Schwinger limit for pair production $E_c=m_{e}^2c^3/\hbar e \simeq 1.3\times 10^{16}$ V/cm, where $m_{e}$ is the electron mass, $e$ is its charge, $c$ is the speed of light, and $\hbar$ is the reduced Planck constant. The field strength depends on the surface density gradient \cite{2010JPhG...37g5201M} and may reach values of $5\times10^{17}$ V/cm \cite{1986ApJ...310..261A,1995PhRvD..51.1440K}. At low temperatures Schwigner mechanism of pair production does not operate due to Pauli blocking of electronic states.
Usov \cite{1998PhRvL..80..230U} have shown that electrosphere of newly formed hot quark star produces powerful electron-positron wind. Recently we reconsidered \cite{2024ApJ...963..149P} the mechanism of pair production in electrosphere of quark stars.

It is crucial that in order to maintain high luminosity in pairs the surface temperature of bare strange star should remain high on a long timescale. For a gamma-ray bursts this is at least a few seconds. In the literature several models for the cooling of newborn strange stars were proposed, and it was argued that they can preserve the surface temperature about $10^{10}$ K for a few seconds \cite{2004csqn.conf..399B, 2021ApJ...922..214L}. Both these works use an integral approach with an average temperature assigned to a given volume of the strange star. At the same time, different approach was proposed in \cite{2002PhRvL..89m1101P} based on the heat transfer equation. The light curves were computed under the condition of neutrino transparency for a period of time starting from one second to $10^{10}$ seconds after strange star formation.

In this paper we reexamine the thermal evolution of strange stars and produce light curves starting at much shorter timescales, motivated by consideration of the Schwinger process. We show that in the case of free neutrino escape the surface temperature of strange star decreases by orders of magnitude on the timescales much shorter than a second. Moreover, we show that even if neutrino trapping is taken into account, the surface of the star cools down very fast due to the Schwinger process and insufficient quark thermal conductivity. Therefore, strange star cannot maintain extreme luminosity on timescales of fraction of a second.

The paper is organized as follows. In Section \ref{hydroequil} we discuss an equation of state, hydrostatic state the heat transfer in hot quark star. In Section \ref{electroequil} we discuss electrostatic configuration and the physical conditions in the electrosphere. In Section \ref{neutrinotransport} thermal evolution of newborn strange star is considered in both transparent and opaque regimes for neutrinos. Discussion and conclusion follow in the last Section.

\section{Hydrostatic equilibrium configuration}
\label{hydroequil}

The conditions of chemical equilibrium and electric neutrality read
\begin{gather}
\mu_s = \mu_u + \mu_e,~~~\mu_d = \mu_u + \mu_e,\\
\frac{2}{3}n_u - \frac{1}{3}n_d - \frac{1}{3}n_s - n_e=0,
\end{gather}
where chemical potential $\mu_i$ and particle density $n_i$ of each specie $i$ (u-,d-,s- quarks and electrons) of completely degenerate matter is
\begin{gather}
n_i = \frac{g_i}{6\pi^2 c^3 \hbar^3}\left( \mu_i^2 - m_i^2 c^4 \right)^{3/2},
\end{gather}
and  phase space factor $g_i$ equals $2$ for electrons and $6$ for quarks.

We assume that $m_e \sim m_u \sim m_d \ll m_s$ and $m_s c^2 = 150$ MeV. Then electric neutrality condition becomes
\begin{gather}
2(\mu_s - \mu_e)^3 - \mu_s^3 - \left( \mu_s^2 - m_s^2 c^4 \right)^{3/2} -\mu_e^3=0.
\end{gather}
We choose  $\mu_s$ as independent parameter and express electron chemical potential $\mu_e$ by solving the cubic algebraic equation
\begin{gather}
\mu_e = \frac{2}{3}\mu_s + 2^{1/3}\frac{2\mu_s^2}{3a}-2^{-1/3}\frac{a}{3}, \\ \notag
a \equiv \left( 20\mu_s^3-9 b+3\left(48\mu_s^6-40b\mu_s^3+9 b^2  \right)^{1/2} \right)^{1/3}, \\ \notag
b \equiv \mu_s^3- \left( \mu_s^2-m_s^2c^4 \right)^{3/2}.
\end{gather}
According to the bag model of non-interacting quark matter ( i.e. neglecting QCD corrections and color superconductivity \cite{FahriJaffe1984, 1995PhRvD..51.1440K})  total pressure $P$ and total energy density $\epsilon$ are
\begin{gather}
P = \sum_{i=u,d,s,e} P_i - B, \\
\epsilon = \sum_{i=u,d,s,e} \epsilon_i +B,
\end{gather}
where $B$ is the bag constant, pressure $P_i$ and energy density $\epsilon_i$ of complete degenerate matter are
\begin{gather}
P_i = \frac{g_i }{48 \pi^2 c^3 \hbar^3} 
\left[  \mu_i \left( \mu_i^2 - m_i^2 c^4 \right)^{1/2} \left(  2 \mu_i^2 - 5m_i^2 c^4 \right) + \right.\\ \notag
\left.m_i^4 c^8 \ln \left( \frac{\mu_i + \sqrt{\mu_i^2 - m_i^2 c^4}}{m_i c^2}  \right)  \right],   \\
\epsilon_i = \frac{g_i}{16\pi^2c^3\hbar^3}\left [  \mu_i \left( \mu_i^2 - m_i^2 c^4 \right)^{1/2} \left(  2 \mu_i^2 - m_i^2 c^4 \right) - \right.\\ \notag
\left. m_i^4 c^8 \ln \left( \frac{\mu_i + \sqrt{\mu_i^2 - m_i^2 c^4}}{m_i c^2}  \right)  \right] .
\end{gather}

On the surface of quark star total pressure $P$ equals zero. As pressure $P_i$ is a function of only one parameter $\mu_s$ one can find the chemical potential $\mu_s$ for a given value of the bag constant $B$.  

 The lowest value of the bag constant is $57$ MeV$/$fm$^3$, which is
determined by the fact that the energy per baryon number of u-,d-quark matter should be higher than the one of $^{56}$Fe. This value of the bag constant leads to the following chemical potentials: $\mu_s = \mu_d \approx 300$ MeV, $\mu_u = \mu_s-\mu_e \approx 281.3$ MeV and $\mu_e \approx 18.7$ MeV.

\begin{figure}[h]
\includegraphics[width=\columnwidth]{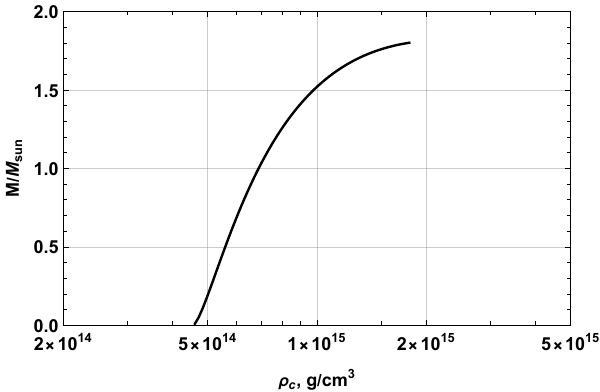}
\caption{Total mass vs. central density for stable quark stars.}\label{fig1}
\end{figure}

\begin{figure}[h]
\includegraphics[width=\columnwidth]{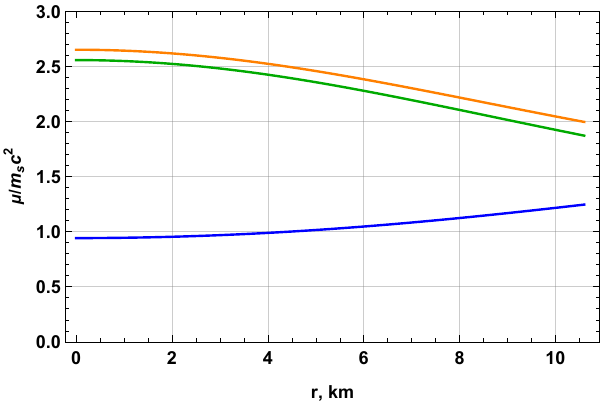}
\caption{Chemical potentials as functions of radius of the star with $M=1.4M_\odot$: $\mu_s=\mu_d$ (orange), $\mu_u$ (green), $10\mu_e$ (blue).}\label{fig2}
\end{figure}

The mass-central density relation is presented in figure \ref{fig1}, with maxumum mass of $1.8\,M_\odot$. In what follows we consider a stellar configuration with the mass $M=1.4 M_\odot$. Chemical potentials as a function of radius for such a star are presented in figure \ref{fig2}. As can be seen, the distribution of quarks is nearly homogeneous.

In this paper we consider SQM without color superconductivity and focus on the surface properties of the star. We calculated average quark chemical potential on the surface $\mu = (\mu_u+\mu_d+\mu_s)/3 = 293.8$ MeV. This value determines the surface chemical potential of electrons and as a result the maximal surface electric field.

Including QCD corrections in the equation of state can only slightly increase the value of $\mu$ \cite{Alford_2005}. Another possibility is connected with the surface tension of quark matter. As shown in \cite{2005ApJ...620..915U} the surface electric field can be 10 times larger than the surface electric field calculated for SQM in the unpaired phase without the surface effects. 

We can omit the last effect, because as we show in section \ref{electroequil} the Schwinger rate contains the Pauli blocking. Therefore, increasing the chemical potential will increase the electric field and enhance Pauli blocking effect. Thus, the resulting Schwinger rate remains almost unchanged.     

As we show in section \ref{neutrinotransport} the key ingredient is thermal conductivity of quark matter, because it affects the temperature gradient on the surface induced by huge pair luminosity. 

Finally, the value of $\mu$ is almost unaffected by the temperature \cite{1995PhRvD..51.1440K}. So, we keep it constant during thermal evolution of the surface.

\section{Electrostatic configuration and Schwinger luminosity}
\label{electroequil}

In this Section, we describe the Schwinger process on the stellar surface, following \cite{2024ApJ...963..149P}. As initial surface temperature is very high, the electric field binding electrons with quarks can produce pairs via the Schwinger process.  Spatial distribution of electron density $n_e$ can be found from the Poisson equation, as described below. As it was pointed out in \cite{2024ApJ...963..149P}, at finite temperature electrostatic configuration is not self-consistent, due to thermal evaporation of electrons. The corresponding stationary regime in hot dynamical electrosphere can be found by solving the Vlasov-Maxwell system.

In this Section we introduce the Compton wavelength of electron $\lambda_e = \hbar/(m_e c)$ and use the following dimensionless quantities: $\tilde \mu_e = \mu_e/(m_e c^2)$ for chemical potential, $\tilde{p} =p/(m_ec)$ for particle momentum, $\tilde T = k_B T/(m_e c^2)$ for temperature, $\tilde{t}=t \lambda_e/c$ for time, $\tilde{z}=z/\lambda_e$ for space coordinate, $\tilde n = n /\lambda_e^{-3}$ for particle density, $\tilde \varphi =\varphi /(E_c\lambda_e)$ for electrostatic potential, $\tilde{E}=E/E_c$ for electric field and $\tilde L=L\lambda_e/(m_ec^3)$ for luminosity. 

Electrons in equilibrium obey the Fermi-Dirac statistics with their distribution function
\begin{equation}\label{FDdistr}
f_e= \frac{1}{1+\exp{\left[\left(\sqrt{\tilde p^2 + 1}-\tilde\mu_e\right)/ \tilde T\right]}}.
\end{equation}
The chemical equilibrium condition for electrons is
\begin{equation}\label{chemeq}
\tilde\mu_e=\tilde\varphi,
\end{equation}
where $\varphi$ is electrostatic potential. 

Introducing dimensionless degeneracy parameter $\eta\equiv \mu/T$ one can write the number density of electrons as
\begin{gather}\label{nedef}
\tilde n_e = \pi^{-2} \int d\tilde p \tilde p^2 \left( 1+e^{-\eta+\sqrt{\tilde p^2+1}/\tilde T}
  \right)^{-1}.
\end{gather}

For ultrarelativistic electrons (when $\tilde p \gg 1$) this expression reduces to polylog function of the third order
\begin{gather}
\tilde n_e \approx - 2 \pi^{-2} \tilde T^3\text{Li}_{3}(-e^\eta),
\end{gather}
which for high degeneracy (when $\eta \gg 1$) reduces to \cite{1995PhRvD..51.1440K,2005ApJ...620..915U}
\begin{equation}\label{napprox}
    \tilde n_e \approx \frac{\tilde\mu_e^3}{3\pi^2}+\frac{\tilde\mu_e \tilde T^2}{3}.
\end{equation}

Poisson equation for quarks and electrons near the surface gives \cite{1986ApJ...310..261A,1995PhRvD..51.1440K}
\begin{equation}\label{poissoneq}
    \frac{d^2\tilde\varphi}{d\tilde z^2}=4\pi \alpha \left[ \tilde n_q - \tilde n_e(\tilde\varphi) \right],
\end{equation}
where $\alpha=e^2/(\hbar c)$ is the fine structure constant and we assume that quark density $n_q$ is constant inside the surface and equals zero outside the surface. The value $n_q$ is given by expression \eqref{nedef} with a fixed chemical potential $\mu_q$ and temperature $T$. In what follows we take $\mu_q=18.7$ MeV. The temperature $T$ is constant inside and outside the surface for both quarks and electrons. 

\begin{figure}[h]
\includegraphics[width=\columnwidth]{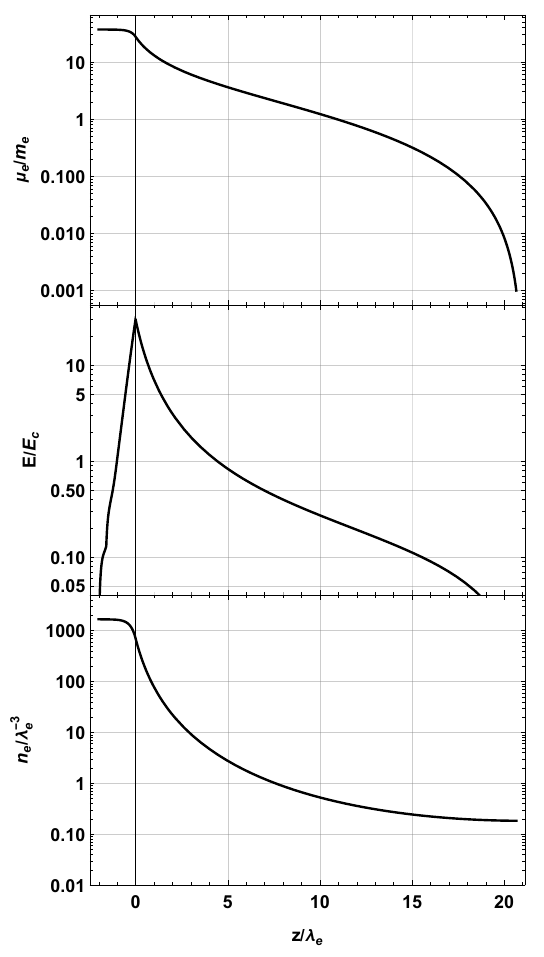}
\caption{Exact solution of Poisson equation \eqref{poissoneq} for $k_BT=m_ec^2$: as function of distance we show chemical potential (top), electric field (middle) and particle density (bottom). Since the chemical potential decreases to zero at $z\simeq21\lambda_e$ the electric field vanishes at this distance. The density of electrons does not vanish here.}\label{fig3}
\end{figure}

The Poisson equation \eqref{poissoneq} is exactly solvable and we present its solution inside and outside the surface $z=0$ for $\tilde T= 1$ in figure \ref{fig3}. Note, that both chemical potential and electric field tend to zero at a finite distance from the surface $z_0$, where electron density has a finite value. This is because the temperature $T$ is constant everywhere. If the asymptotic formula \eqref{napprox}, is used instead, electron density tends to zero together with the chemical potential. Clearly both approximations fail at large distances from the surface. Solution valid for large distances can be obtained only by direct solution of the Vlasov-Maxwell equation. 

To describe pair creation for non-vacuum initial state it is necessary to take into account Pauli blocking effect \cite{2023PhRvD.108a3002P,2023PhRvE.107c5204B}. We use the differential pair creation rate given in \cite{1987PhRvD..36..114G} with additional Pauli blocking factor $(1-f_e)$. The rate is then given by an integral over particle momentum $d^3p_e$ in the following form
\begin{gather}
\label{ndotschwinger}
\frac{d\tilde n_\pm}{d\tilde t} = - \frac{\tilde E}{2\pi^2} \int\limits_0^\infty d\tilde p \tilde p \left(1-f_e\right) \\ \notag \times\ln\left[  1-\exp{\left(-\frac{\pi(\tilde p^2+1)}{\tilde E} \right)}  \right]. 
\end{gather}

Finally, we estimate pair luminosity as
\begin{gather}\label{Lschwinger}
\tilde L_\text{Schwinger} = 4\pi \tilde R^2 \gamma \int\limits_0^{z_0} \frac{d\tilde n_\pm}{d\tilde t} d\tilde z,
\end{gather}
where $\gamma$ is the Lorentz factor of positrons. 

\begin{figure}[h]
\includegraphics[width=\columnwidth]{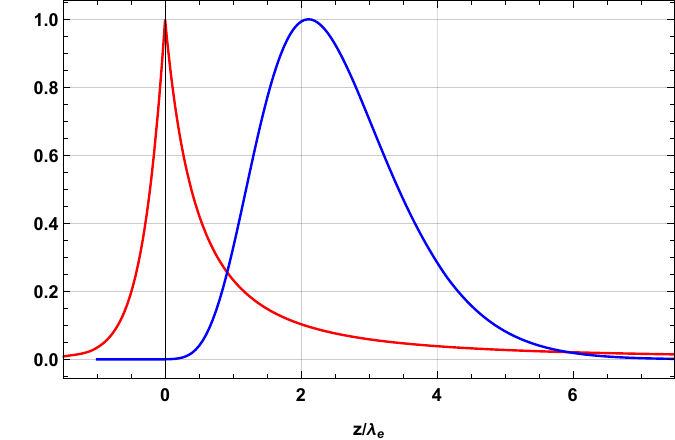}
\caption{Normalized pair rate \eqref{ndotschwinger} (blue) with the corresponding normalized electric field (red) for $k_BT=m_ec^2$.}
\label{fig4}
\end{figure}

To obtain luminosity in pairs we use the distribution function \eqref{FDdistr} in equation \eqref{ndotschwinger} where the chemical potential is given by the solution of the Poisson equation \eqref{poissoneq} for a given constant temperature. As a result, $\dot n_\pm$ is a peaked function of $z-$coordinate. At low temperatures $T<5 m_e c^2$ the peak is localized outside the electrosphere, see figure \ref{fig4}. This effect is caused by Pauli blocking, incorporated in the factor $(1-f_e)$ in equation \eqref{ndotschwinger}.

Here we assume pairs are created at rest in the position of the peak of the rate, see figure \ref{fig4}, and compute $\gamma$ from the relativistic equation of motion $d\sqrt{\gamma^2-1}/d\tilde t=\tilde E$. While positrons are accelerated in the electrosphere, electron evolution is more complicated: after evaporation part of them returns on the surface being attracted by the electric field, and thermalizes again. The total pair luminosity is well approximated by \eqref{Lschwinger} since even if electrons have lower luminosity than positrons, eventually positrons share their kinetic energy with electrons.

For arbitrary distribution functions of electrons $f_-$ and positrons $f_+$ one can define the pair luminosity of the outflow at a distance $\tilde z$ from the boundary as
\begin{equation}
\tilde L_\pm = 4\pi \tilde R^2 \int 2\frac{d^3\tilde p}{(2\pi)^3} \tilde p_{||} f_\pm(\tilde z).
\label{Lpair}
\end{equation}

Unlike the estimate \eqref{Lschwinger}, equation \eqref{Lpair} represents the flux of particles through a given boundary $\tilde R+\tilde z$.

The set of Vlasov-Maxwell equations for distribution functions $f_\pm$ and electric field $\tilde E$ in our case reduces to the Vlasov-Amp\'ere system \cite{2024ApJ...963..149P}
\begin{gather}
\label{vlasovampere}
\frac{\partial f_\pm}{\partial \tilde t} +\frac{\tilde p_{||}}{\tilde p^0}\frac{\partial f_\pm}{\partial \tilde z}\mp \tilde E\frac{\partial f_\pm}{\partial \tilde p_{||}}= \\ \notag
-(1-f_{+}-f_{-})|\tilde E|\text{ln}\left[1-\exp\left(  -\frac{\pi(1+\tilde p_\perp^2)}{|E|}\right)\right] \delta(\tilde p_{||}),   \\ \notag
\frac{\partial \tilde E}{\partial \tilde t} = 2 \alpha\int d^3 \tilde p \frac{\tilde p}{\tilde p^0}(f_- - f_+)+ 
4 \alpha\frac{|\tilde E|}{\tilde E}\int d^3\tilde p \tilde p_0 \times  
\\ \notag
(1-f_{+}-f_{-})\text{ln}\left[  1-\exp\left(-\frac{\pi(1+\tilde p_\perp^2)}{|\tilde E|}\right)  \right]\delta(\tilde p_{||}),
\end{gather}
where $\tilde p_0 = \sqrt{\tilde p_{||}^2+\tilde p_\perp^2+1}$.

\begin{figure}[h]
\includegraphics[width=\columnwidth]{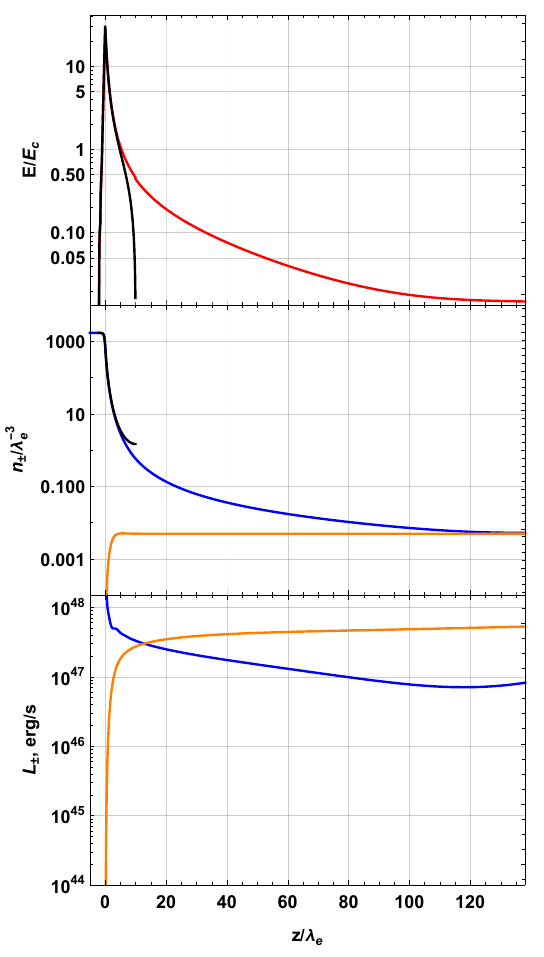}
\caption{Stationary solution of Vlasov-Maxwell equations for electrosphere for $k_BT=2m_ec^2$ with dependence on distance of electric field (top), particle density (middle) and luminosity (bottom). Approximate electrostatic solution given by eq. (\eqref{poissoneq}) is shown in black. Colors correspond to: electric field (red), electrons (blue) positrons (orange), and electrostatic solution (black).}
\label{fig3a}
\end{figure}

The stationary numerical solution (obtained for large enough time) of these equations is shown in figure \ref{fig3a}.
This solution is compared to the electrostatic solution of equation \eqref{poissoneq} for constant temperature. There is a good agreement in the region of overcritical electric field, where pairs are produced. However, electrons are redistributed to much larger distances, bringing with them electric field. As a consequence, positrons are accelerated to larger energies and have higher luminosity. At large distance $z>100$ pair outflow becomes electrically neutral. In this example luminosity in positrons at $z=140\lambda_c$ is $5\times 10^{47}$ erg/s, luminosity in electrons is $8\times 10^{46}$ erg/s, while pair luminosity estimated from \eqref{Lschwinger} is $4\times 10^{47}$ erg/s.


\section{Heat transfer and neutrino losses}
\label{neutrinotransport}

So far we focused on luminosity in pairs as a function of temperature at the surface of quark star, but did not consider the evolution of this surface temperature. Indeed, we implicitly assumed that the temperature can be maintained for a time interval long enough to establish the stationary solution shown in figure \ref{fig3a}. 

Now we are going to consider heat transfer in the quark star accounting for temperature evolution on much large timescales than the ones studied above. First we treat neutrino transparent case, following \cite{2002PhRvL..89m1101P, 2001A&A...368..561B, 2022A&A...663A..19Z, PhysRevC.85.035805}.

Specific heat capacity $c_v$ for unpaired degenerate quarks is \cite{IWAMOTO19821}
\begin{gather}
c_v = \pi^2 n \frac{k_B^2 T}{\varepsilon_F} =  
\frac{8\pi^3 k_B^2}{3 c^3 h^3}\mu^2 T.
\end{gather}
 Heat conductivity $\kappa$ for unpaired degenerate quarks is \cite{Haensel:1989ja, 1991NuPhS..24...23H}
\begin{gather}
\kappa = \frac{3}{\pi^3}\sqrt{\frac{3}{2\pi}}\frac{\hbar k_F^3 c^2}{\alpha_s^{1/2}T}, 
\end{gather} 
where $k_F$ is Fermi wavenumber of quarks,  $\alpha_s = 0.3$ is QCD coupling constant.

In unpaired quark matter neutrino
emissivity is dominated by the direct URCA processes of light quarks $d\rightarrow u + e + \bar\nu_e$ and $u+e\rightarrow d+\nu_e$. The energy emission per unit time and volume is  \cite{IWAMOTO19821}  
\begin{gather}
    Q_\nu = 3\times10^{25}\alpha_s \frac{\mu_u\mu_d}{(400\text{MeV})^2}\frac{\mu_e}{10\text{MeV}}T^6_9.
\end{gather}

The temperature evolution is determined from the heat transfer equation with boundary conditions
\begin{gather}\label{heateq}
c_v \frac{\partial T}{\partial t} = - \frac{1}{4\pi r^2} \frac{\partial (4\pi r^2 F_r)}{\partial r} - Q_\nu,~~~F_r = -\kappa \frac{\partial T}{\partial r},  \\
\frac{\partial T(t,r=R)}{\partial r}= - \frac{F_{pair}}{\kappa} =- \frac{L_{pair}}{4\pi R^2 \kappa}, \\ \notag
T(t=0,r)=T_0,
\end{gather}
here $L_{pair}$ is the luminosity in electron-positron pairs (referred to as Schwinger luminosity). We approximate this luminosity by expressions \eqref{nedef} and \eqref{ndotschwinger} for electrostatic solutions at constant temperature, using equation \eqref{Lschwinger}.

\subsection{Neutrino transparent regime}

In addition to neutrino loss from the entire volume of the star, at the stellar surface $F_r(r=R)$ electron-positron pairs are created via the Schwinger process and contribute to the energy flux.

\begin{figure}[h]
\includegraphics[width=\columnwidth]{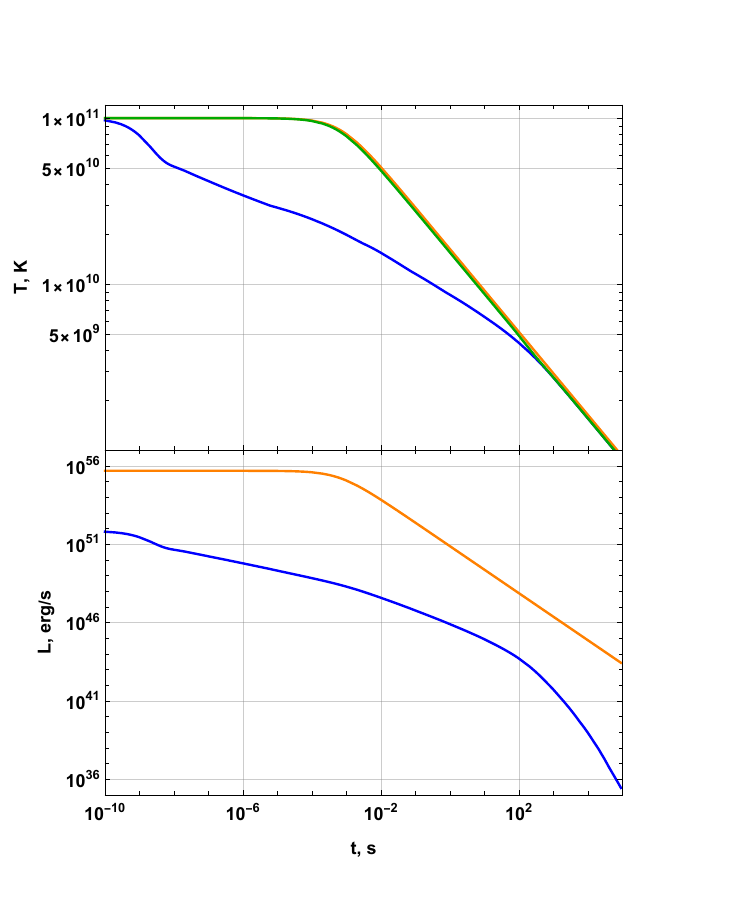}
\caption{Top: Time evolution of the star temperature at selected radii. Blue: surface temperature. Orange: central temperature. Green: temperature at one meter under the surface. \\
Bottom: Time evolution of pair luminosity (blue) and neutrino luminosity (orange).}\label{fig56}
\end{figure}

We solved numerically equation \eqref{heateq} for the stellar configuration with the mass $M=1.4 M_\odot$. The temperature evolution of the star is shown in figure \ref{fig56} together with luminosity evolution.  After the time moment $t>1$ second, our curves for neutrino demonstrate the same shape as in fig. 1 in \cite{2002PhRvL..89m1101P}. 

Note that the heat flux distribution has a highly peaked shape near the surface, which leads to rapid decrease of the surface temperature. Clearly, the cause is the low thermal conductivity of quark matter which is unable to transfer heat from interior of the star to its surface, where it is quickly released due high luminosity of the Schwinger process. One can give the following simple estimate of the temperature gradient: $L_\text{pair}\simeq 10^{52}$ erg/s, $4\pi R^2 \simeq 10^{13}$ cm$^2$, $\kappa\simeq10^{22}$ erg/s/cm/K $\implies\partial T/\partial r \simeq 10^{17}$ K/cm.

The most important result is that the temperature evolution on the surface is independent on interior of the star. Surface temperature decreases up to few $5\times 10^9$ K at $t\simeq10^{2}$ seconds. The corresponding luminosity is about $10^{43}$ erg/s. The temperature gradient disappears only at $10^2$ seconds, when the pair luminosity starts to decrease rapidly.

The total energy emitted in neutrinos is about $10^{53}$ ergs, while the total energy emitted in pairs is only $8\times 10^{46}$ ergs.

In the next Section we argue that rapid surface temperature decrease occurs also in neutrino trapped regime.


\subsection{Neutrino opaque regime}

It is believed that neutrinos cannot escape freely immediately after strange star formation, until internal temperature decreases below $10^{10}$ K \cite{2002PhRvL..89m1101P}. Following that idea, the integral models for neutrinosphere was developed in \cite{2004csqn.conf..399B,2021ApJ...922..214L}, where quark star is divided into two parts having different temperatures, separated from each other by a neutrinosphere with radius $R_\nu<R$. Neutrinosphere emits blackbody neutrino spectrum from its surface. The results of \cite{2004csqn.conf..399B} imply that neutrino opacity delays the process of cooling by an order of magnitude when compared to the transparent case. The results in \cite{2021ApJ...922..214L} imply that high luminosity in pairs up to $10^{50}$ erg/s can be sustained for long time, up to $10$ seconds. In this Section we compare the integral model developed in \cite{2021ApJ...922..214L} with the differential models adopted in our work.
\begin{figure}[h]
\includegraphics[width=\columnwidth]{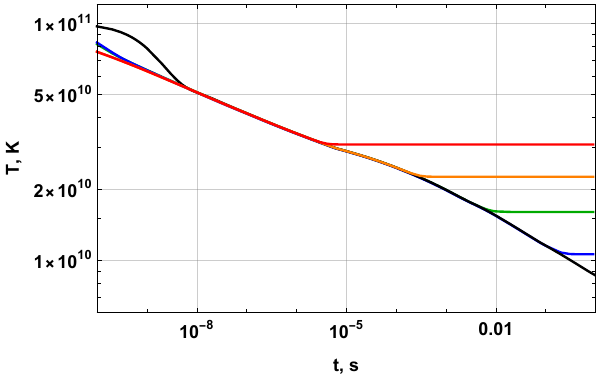}
\caption{Surface temperature evolution for neutrino transparent shell of the star with selected positions of neutrinosphere: $R-R_\nu=0.001$ cm (red), $R-R_\nu=0.01$ cm (orange), $R-R_\nu=0.1$ cm (green), $R-R_\nu=1$ cm (blue). Black curve represents blue curve from figure \ref{fig56} (top) for convenience.}
\label{fig7}
\end{figure}

Our simulations for neutrino transparent regime show that the surface temperature decreases very fast due to high luminosity of the Schwinger process. In order to reveal the role of neutrinosphere one can perform the following simple test.

\begin{figure}[h]
\includegraphics[width=\columnwidth]{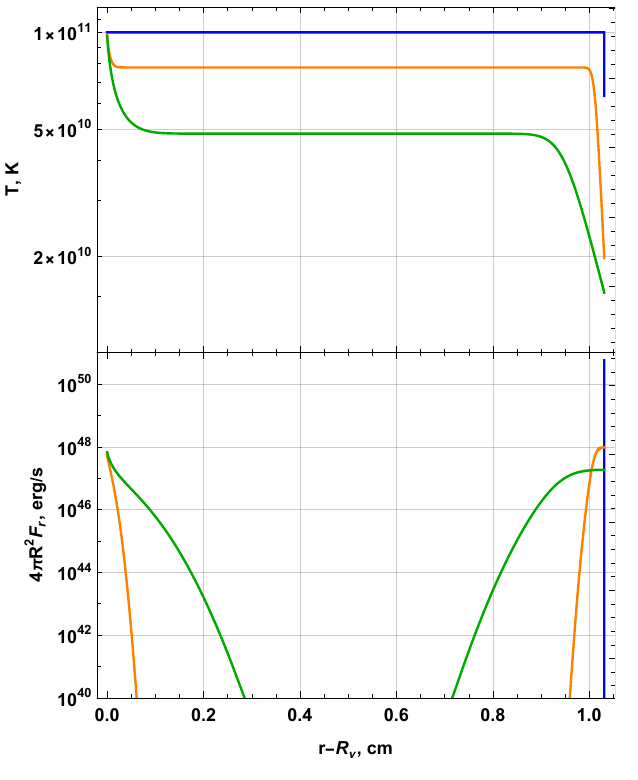}
\caption{Temperature distribution (top) and heat flux distribution (bottom) inside the shell for selected time moments: $10^{-10}$ s (blue); $10^{-3}$ s (orange); $10^{-2}$ s (green). The case of neutrino transparent shell of the star for the position of neutrinosphere $R-R_\nu\approx 1$ cm.}\label{fig8}
\end{figure}

Consider an outer shell of the star and fix the temperature at its inner boundary to be equal to the initial temperature $T(t,r=R_\nu)=T(t=0,r)=T_0$. We solve numerically the heat transfer equation for different thickness of the shell $R-R_\nu=\{0.001,0.01,0.1,1\}$ cm. The results for the surface temperature are presented in figure \ref{fig7}. It is clear that the surface temperature decreases very fast in all cases. When the thickness increases up to $R-R_\nu=1$ cm cooling occurs in the same way as for neutrino transparent star, that is the surface temperature is about $10^{10}$ K at the time moment $t\simeq 0.2$ seconds. Finally, in figure \ref{fig8} we show temperature distribution $T(r)$ and heat flux distribution $F_r(r)$ (for convenience we plot the value $4\pi R^2 F_r$) inside the shell with thickness about 1 cm at different instants of time. Clearly, initially very steep temperature gradient emerges instantly and remains localized at the surface, due to weak thermal conductivity of quarks.

In realistic model neutrinosphere moves toward the center of the star and at the same time loses its temperature. It means that our model overestimates the surface temperature. Thus, our results imply that the surface temperature of a newborn strange star made of  unpaired quarks decreases below the value $10^{10}$ K on the timescales $\Delta t\ll1$ second, and the reason for that is the very high luminosity due to the Schwinger process and weak thermal conductivity of unpaired quarks.

\section{Discussion and conclusions}

In this paper thermal evolution of a newborn strange star  made of unpaired strange quark matter is considered, taking into account kinetics of pairs created in its electrosphere. Numerical solution of the heat transfer equation in neutrino transparent regime shows that the surface temperature of newborn strange star reduces from initial $10^{11}$ K to $10^{10}$ K on a very short timescale $\Delta t \sim 0.1$ seconds. At the same time, interior temperature starts to change on longer timescale, about $10^{-3}$ seconds, due to the URCA process. The reason of high temperature gradient, established near the surface is a combination of high power of the the Schwinger process and low thermal conductivity of  unpaired quarks. Hence the surface of newborn strange star cannot sustain high luminosity on long timescales, as required for the explanation of extreme astrophysical transient events, such as gamma-ray bursts.

Here we did not consider subsequent evolution of pair outflow at large distances from the surface of the star, as this topic is already covered in the literature \cite{2004ApJ...609..363A,2005ApJ...632..567A}. Note that the luminosity in pairs decreases from $10^{46}$ erg/s at $t\sim 1$ second to $10^{36}$ erg/s at $t\sim10^5$ seconds. This is precisely the luminosity range where a transition from black body-like spectrum (from thermalization of pairs due to large optical depth of the outflow) to annihilation line occurs \cite{2004ApJ...609..363A}. This effect is potentially observable and may represent the unique feature of quark star cooling.

We also considered the effects of neutrino trapping at high temperatures and found that simplified integral models, such as the ones presented bin\cite{2004csqn.conf..399B,2021ApJ...922..214L} do not account for the sharp temperature gradient at the surface of quark star. They fail to estimate the surface temperature and therefore cannot be used to obtain pair luminosity as a function of time. To our knowledge, detailed calculations of the dynamics of neutrinosphere inside a newborn strange star is not performed yet. We are developing a spherical Boltzmann code for neutrino transfer (\cite{2017rkt..book.....V}) in order to close this gap. Nevertheless, our toy model for neutrino transparent shell, described in section above, clearly shows that the conclusion obtained for the neutrino transparent case holds. 

Note, that a true QCD phase of strange quark matter is still uncertain. If color superconductivity takes place, it can increase thermal conductivity of quarks \cite{RevModPhys.80.1455}.

Thermal conductivity in the two-color–flavor superconductor (2SC) phase was revised in \cite{PhysRevC.90.055205}. It was shown that at high temperatures quark-quark interactions do not contribute to the total thermal conductivity and electron-quark interactions become the main ones. The corresponding thermal conductivity $\kappa_e$ for $T=10^{11}$ K and $\mu_q=300$ MeV is of the order of $10^{21}$ erg/s/cm/K. Comparing the last value with the unpaired case studied in this paper $\kappa \simeq 10^{22}$ erg/s/cm/K we see that the strange star surface made of 2SC quarks can not preserve high temperature for a long time. 

According to \cite{PhysRevC.66.015802} thermal conductivity of the color-flavor-locked (CFL) phase can be of the order of $10^{33}$ erg/s/cm/K. Strange star in CFL phase does not contain electrons, but the surface electron layer can appear, because of the number of massive quarks (s quarks) is reduced near the boundary relative to the number of massless quarks. In this case the structure of electrosphere  differs significantly \cite{PhysRevD.70.067301}.

We will consider strange quark stars with color superconductivity in the next paper.



{\bf Acknowledgements.} 
We thank the anonymous referee for his comments which improved the paper.
This work is supported within the joint BRFFR-ICRANet-2025 funding programme under the grant No. F25ICR-001.

\bibliography{total}{}
\bibliographystyle{unsrt}


\end{document}